\documentclass{vldbhacked}
\usepackage[dvips]{epsfig}
\usepackage{color}
\usepackage{hyperref}

\newtheorem{algorithm}{Algorithm}

\title{Search and Navigation in Relational Databases}

\author{Richard Wheeldon, Mark Levene and Kevin Keenoy}
\institution{
	School of Computer Science and Information Systems\\
  Birkbeck University of London\\
  Malet St, London\\
  WC1E 7HX, United Kingdom\\
  \{richard,mark,kevin\}@dcs.bbk.ac.uk
}

\begin{document}

\maketitle

\begin{abstract}
We present a new application for keyword search within relational databases,
which uses a novel algorithm to solve the join discovery problem by finding
Memex-like trails through the graph of foreign key dependencies. It differs
from previous efforts in the algorithms used, in the presentation mechanism
and in the use of primary-key only database queries at query-time to maintain
a fast response for users. We present examples using the DBLP data set.

\end{abstract}

\section{Introduction}
\label{sec:intro}

\begin{quote}
	``Future users of large data banks must be protected from having to know how the
	data is organized in the machine (the internal representation).''
	\flushright{E. F. Codd \cite{CODD70}}
\end{quote}

We consider that for many users of modern systems, being protected from the
internal structures of pointers and hashes is insufficient. They also need to
be spared the requirement of knowing the logical structures of a company or of
its databases. For example, customers searching for information on a particular
product should not be expected to know the address at which the relevant data is
held. But neither should they be expected to know part numbers or table names in
order to access this data, as required when using SQL.

Much of today's corporate data resides in relational databases,
comprising a large chunk of what is known as the ``hidden'' or ``deep'' web.
The word ``hidden'' means that, from a practical point of
view, this data is hidden from conventional search engines;
the word ``deep'' is intended for greater accuracy, meaning that the 
data can only be accessed through a specialised query interface.
It is estimated that the deep web contains 500 times more information 
than is visible to conventional search engines \cite{BERG00}.

One way for users to access data in the deep web is through a site-specific
search engine, such as the query interface at \href{http://www.amazon.com}{Amazon.com}.
One can imagine that Amazon have a relational database storing all their
catalogue information, over which the full-text query facility was developed.
Research shows that users actively use such interfaces and expect major
web sites to support unstructured search facilities \cite{NIEL97}. These
interfaces are more natural that the SQL syntax supported directly by the
database. However, the full-text search will result in a loss of expressiveness
relative to the full expressive power of SQL, which is an issue that we will
partially explore. One can argue that many end users do not need access to
the full expressive power of SQL. Studies of keyword-based search engines on
the web have shown that users type short queries, rarely use advanced features
and are typically bad at query reformulation \cite{SILV99,JANS98,SPIN98,SPIN02}. It is
likely that profiles for users of a database search facilities will
reveal similar behaviour.

Vannevar Bush's seminal 1945 paper ``As We May Think'' first suggested the
concept of a trail as a sequence of connected pages, with a future machine
called Memex which would help the user build a ``web of trails'' \cite{BUSH45}.
The concept of trails is well established in the hypertext community \cite{REIC99}.
These sequences of pages have also been referred to as tours or paths and
several hypertext systems have allowed for their construction, but no previous
system has allowed the automated construction of these trails or allowed the
construction of trails across tables in relational databases.

We have previously developed a tool which automates trail discovery
for providing users with navigational assistance and search facilities whilst
surfing a web site. We first introduced the system in \cite{LEVE01a} and
introduced a new graph-based interface alongside the work with automated
Javadoc documentation in \cite{WHEE02}. The navigation engine works by
finding \emph{trails} - sequences of linked pages, which are relevant to
the user query. These trails are presented to the user in a tree-like
structure with which they can interact. User studies have shown the value
of providing contextual information in our combined search and navigation
interface \cite{MATH01}. In a serious of tasks related to the UCL web site,
users found the information they were looking in less time, with fewer clicks,
and with a higher degree of satisfaction compared with using the Google
index or UCL's own Compass system (which has subsequently been replaced).

Building on this work with hyperlinked web pages, we have developed a tool
called \emph{DbSurfer}
which provides an interface for extracting data from relational databases. This
data is extracted in the form of an inverted index and a graph, which can together
be used to construct trails of information, allowing free text search on the
contents. The free text search and database navigation facilities can be used
directly, or can be used as the foundation for a customized interface. We hope
that the trail structure and interface provided will provide the same benefits
for users of the database search and assisted navigation facilities as for
users of the web site interface.

The rest of the paper is organized as follows. In section~\ref{sec:indexing}
we describe our methods of indexing the content of a relational database for
keyword search. In section~\ref{sec:trails} we describe the algorithms for
extending this to compute joins by building trails.
Section~\ref{sec:arch} gives an overview of the system's architecture.
Section~\ref{sec:xml} discusses the work done to incorporate XML indexing into
the system and section~\ref{sec:query} discusses how this compliments the
preceeding work to provide expressive queries for solving user's information needs.
Section~\ref{sec:examples} gives examples of this technique using DBLP data.
Section~\ref{sec:evaluation} gives an overview of preliminary work into the
evaluation of DbSurfer and related systems.
Section~\ref{sec:related} discusses related work and in section~\ref{sec:future}
we discuss directions for future research.

\section{Indexing a Relational Database}
\label{sec:indexing}

\subsection{From Relations to a Full-Text Index}

A single relation (or table) is a set of rows each of which can be addressed
by some primary key. To index these rows we extract the data from each row in
turn and construct a \emph{virtual document} or web page, which is indexed by our
parser. This parser will recognize web content and handle document formats such
as Postscript, PDF, Microsoft Office, Shockwave Flash and RPM package formats
which may be stored as binary objects in a database, but over which indexes may
never be created. The textual content of this document is extracted and stored
in an inverted file \cite{HARM92}. During the parsing stage, URLs are retrieved
which reference other web pages. These URLs may be sent to our crawler and the
pages added to the same index. The inverted file is indexed such that the
posting lists contain normalized \emph{tf.idf} entries as prescibed in
\cite{SALT98} although many variations are possible \cite{BAEZ99}.

Whilst these virtual documents are transient and exist only for the time
it takes to be indexed, the entries in the posting lists provide references
to a servlet which will reproduce a customized page for each row entry. This
is achieved by extracting the data, converting it to XML using a SAX generator,
and applying an XSLT stylesheet to the resulting page \cite{HARO01}. Binary
data is handled with a separate servlet accessed via links from these pages.
The data for these pages is always accessed via a primary key, so the page
display is almost instantaneous. This is essential for providing the quick
responses that users insist on \cite{NIEL00}. It is a practical impossibility
to guarantee response times on large databases when queries may contain full
table scans and much work goes into avoiding them in traditional e-commerce
systems \cite{GURR96}.

The primary key may not be a convenient index to embed in a url format. For
example, it may be a composite key with a large number of attributes or even
a binary object (BLOB). To cover these possibilities and make the system robust
we create a second identifier which identifies this key, giving a two step
lookup process. This index is held externally to the database.
Oracle databases contain a unique rowid for each table which we can
index, saving us from this two-stage process. Similar optimizations
exist for other databases, but these have yet to be fully exploited.

\subsection{Generating the Link Graph}

Answers to user's queries may not be contained in a single table. Often the
results are spread over several tables which must be joined together. We
can answer such queries with the help of a \emph{link graph}.
We have shown how we can create an inverted file containing URLs, some of
which reference traditional web pages and some of which reference servlets
which return customized views of database content. All these URLs are assigned
a separate 32-bit number which identifies them. It is these numbers which are
stored in the inverted file, and it is these numbers which are stored in
the link graph.

The link graph is constructed by examining the foreign key constraint of the
database (either by accessing the data dictionary table or via the JDBC
APIs) and the data entries themselves. Each matching set of ($table$, $attribute$)
pairs where there is a recognized referential constraint generates a
bi-directional link.
Each row entry is converted to a URL and the indexes for these URLs are
added to the link graph. The set of links between web pages and between
database rows and web pages is also added to this graph. The approach is
equivalent to the Link1 database presented in \cite{RAND01} and the same
techniques for improving the memory usage characteristics should work equally
well in this case. These techniques are not used in our DBLP demo as the DBLP
example is sufficiently small to be easily contained without compression,
and the increased query time due to the cost of compression would be an
unneccessary sacrifice.
The strength of this approach is that it allows transparent access to the
database in a manner which is compatible with access to any other web page
on the web site and for relational data to be joined with relevant web data.

\section{Computing Joins with Trails}
\label{sec:trails}

Given the graph of related elements, we can utilise our \emph{navigation
engine} approach to construct join sequences as trails. The navigation engine
works in 4 stages. The first stage is to calculate scores for each of the
nodes matching one or more of the keywords in the query, and isolate a small
number of these for future expansion. The second stage is to construct the trails
using the \emph{Best Trail} algorithm \cite{WHEE03f}. The third stage involves
filtering the trails to remove redundant information. In the fourth and final
stage, the navigation engine computes small summaries of each page or
row and formats the results for display in a web browser.

Each node (whether database row or web page) is scored using Salton's normalized
$tf.idf$ metric \cite{SALT98}, although other IR metrics can be used. Selection
of the \emph{starting points} is done by combining the $tf.idf$ scores with a
node ranking metric called \emph{potential gain}, which rates the navigation
potential of a node in a graph based upon the number of trails available from
it.

The best trail algorithm takes as input the set of starting nodes and builds a
set of navigation trees, using each starting point as the root node.
Two series of iterations are employed for each tree
using two different methods of probabilistic node selection. Once a sufficient
number of nodes have been expanded, the highest ranked trail from each tree is
selected. The subsequent set of trails is then filtered and sorted.
Figure~\ref{fig:best} shows the algorithm in more detail. In this figure,
$S$ represents the set of starting points, $M$ represents an optional number of
repetitions to be performed, reducing the element of chance in the calculation
and $I_{expand}$ and $I_{converge}$ control the number of expansion and
convergence iterations. $D$ represents the navigation tree, which grows according
to the average cardinality of the records in the database. The maximum size of
$D$ is fixed and adding nodes within $D$ is a trivial operation.
A single leaf node, $t$, of $D$, referred to as a \emph{tip}, is
$select$ed during each iteration. $\rho$ is a function from the set of
trails to the set of real numbers, used to assign scores to the trails for
selection. Two functions have been chosen specifically to allow a $O(Log(|D|))$
selection time. The chosen tip is $expand$ed and the linked nodes are assigned
new tips in $D$. After the expansion and convergence iterations have been completed
the highest ranked trail from each expanded starting point is selected by the
function $best()$ and the resulting trail is added to the set of candidate trails, $B$.
$df$ is a discrimination factor which speeds up the convergence process and forces
behaviour closer to that of a best-first approach. With appropriate choice of
parameters ($I_{expand}=0$, $df \approx 0$), the best trail algorithm can emulate
the simpler best-first algorithm.

\index{Algorithm!Best Trail}
\begin{figure}[htbp]
	\begin{algorithm}[Best\_Trail($S, M, \rho$)]\label{alg:best}
	\begin{rm}
	\begin{tabbing}
	t1\=t2\=t3\=t4\=t5\=t6\= \kill \\
	1.  \> \> {\bf begin} \\
	2.  \> \> \> {\bf foreach} $u \in S$ \\
	3.  \> \> \> \> {\bf for} $i = 0$ to $M$ {\bf do} \\
	4.  \> \> \> \> \> $D \leftarrow \{u\}$; \\
	5.  \> \> \> \> \> {\bf for} $j = 0$ to $I_{explore}$ {\bf do} \\
	6.  \> \> \> \> \> \> $t \leftarrow select(D, \rho)$; \\
	7.  \> \> \> \> \> \> $D \leftarrow expand(D, t)$; \\
	8.  \> \> \> \> \> {\bf end for} \\
	9.  \> \> \> \> \> {\bf for} $j = 0$ to $I_{converge}$ {\bf do} \\
	10. \> \> \> \> \> \> $t \leftarrow select(D, \rho, df, j)$; \\
	11. \> \> \> \> \> \> $D \leftarrow expand(D, t)$; \\
	12. \> \> \> \> \> {\bf end for} \\
	13. \> \> \> \> \> $B \leftarrow B \cup \{best(D)\}$ \\
	14. \> \> \> \> {\bf end for} \\
	15. \> \> \> {\bf end foreach} \\
	16. \> \> \> {\bf return} $B$ \\
	17. \> \> {\bf end.} \\
	\end{tabbing}
	\end{rm}
	\end{algorithm}
	\caption{
		\label{fig:best} The Best Trail Algorithm.
	}
\end{figure}

The trails are scored according to two simple metrics: the sum of the unique
scores of the nodes in the trail divided by the length plus a constant, and the
weighted sum of node scores, where weights are determined by the position in the
trail, and the number of repetitions of that node. Nodes which occur early in the
trail receive a higher weight, whilst nodes which occur later or are repeated
receive a lower weighting. These functions encourage non-trivial trails, whilst
discouraging redundant nodes. Two navigation trees are constructed
from each node, one for each of these functions. All trail ranking is done by
comparing firstly the number of keywords matched in a trail, secondly, the greatest
number of keywords matched by any given node in the trail and finally, the
trail score.

Filtering takes place using a greedy algorithm and removes any sequences of
redundant nodes which may be present in the trail. Redundant nodes are nodes
which are either deemed to be of no relevance to the query or replicate exactly
content found in other nodes. It should be noted that this concept can easily
be extended to include removal of near-duplicates \cite{BROD00}.

Once they have been filtered and sorted, the trails are returned to the user
and presented in our \emph{NavSearch} interface, the two main elements of
which are a {\em navigation tool bar} comprising of a sequence of URLs (the
``best trail'') and a {\em navigation tree window} with the details of all the
trails. The content of any row can be examined by clicking on any likely looking
entry or by examining the summary data in the enhanced tooltips.

\section{Architecture}
\label{sec:arch}

Conventional web search engines usually use an architecture pattern
comprising three components - a robot or crawler, an indexer and a query engine
\cite{PINK94,BRIN98,RISV02}. We also follow this design but
augment the information retrieval engine with our trail finding system.
In addition, we augment the crawler with the database indexer described above.
A key difference betweeen the DbSurfer and a conventional search engine
is that a search engine traditionally returns links to pages which are
logically and physically separated from the pages of the servers performing
the query operations, whereas the links returned by the DbSurfer refer mostly
to the row display servlet we have described.

\begin{figure*}[htbp]
  \begin{center}
		\psfig{figure=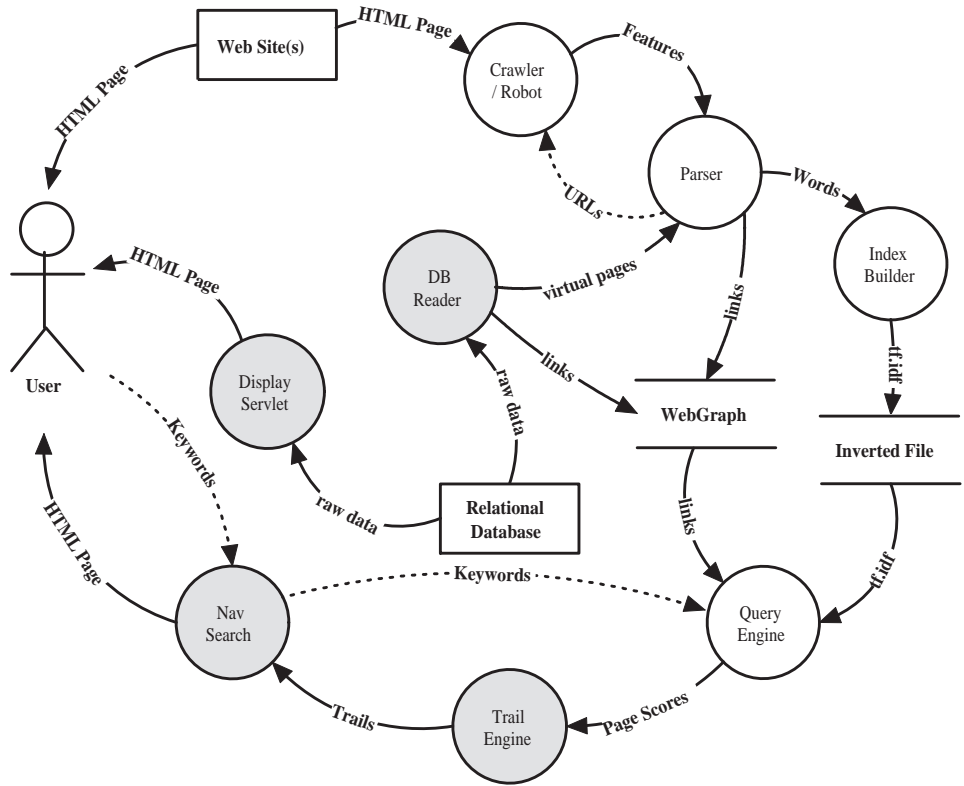, width=375pt}
  	\caption{
			\label{fig:dbsurfer-architecture}
			Architecture of DbSurfer. Closed boxes represent the external sources
			of data which the user is interested in; open boxes represent internal
			data stores; unshaded circles represent processes typically	associated
			with search engines; shaded circles represent processes unique to
			DbSurfer; solid arrows represent data flow and dotted arrows represent
			flows of important information (URLs and Queries). Simple keyed ``get''
			instructions (for example in HTTP requests) are omitted for clarity.
		}
  \end{center}
\end{figure*}

Figure~\ref{fig:dbsurfer-architecture} shows the detailed architecture. The data
from the database is retrieved by the DbReader when the index is built and
by the display servlet when examining the constructed trails.

The database indexer (or reader) works by connecting to the database, selecting
all the accessible tables and views available, and asking the administrator which of
these should be indexed. The program will then extract the referential
constraints for all of the selected tables and build a lookup table. This is
kept separate from the main index and used by both the indexer and the display
servlet.

\section{Semi-Structured Data and XML}
\label{sec:xml}

A relational database can be viewed as a special case of a more general model of
semistructured data and XML \cite{ABIT00}. Hence it might not be suprising that
we can handle XML data using DbSurfer. Indeed that is all DbSurfer does! The
virtual documents alluded to in section~\ref{sec:indexing} are XML
representations of relational tuples. Figure~\ref{fig:dblp-xml} shows an example
of this from a row in the DBLP schema discussed in section~\ref{sec:examples}.
The superfluous \texttt{row}	element has been added for compatibility with the
emerging SQL/XML standard \cite{EISE02}. We note that the proposed standard
includes generation of an XML Schema which we neither construct nor require
at present.

\begin{figure*}[htbp]
  \begin{center}
    \begin{rm}
      \begin{tabbing}
        t1\=t2\=t3\=t4\=t5\=t6\= \kill \\
         1. \> \>  $\langle$PUBLICATION$\rangle$  \\
				 2. \> \> \> $\langle$row$\rangle$ \\
         3. \> \> \> \>  $\langle$JOURNAL$\rangle$ Advances in Computers $\langle$/JOURNAL$\rangle$ \\
         4. \> \> \> \>  $\langle$KEY$\rangle$ journals/ac/Dam66 $\langle$/KEY$\rangle$  \\
         5. \> \> \> \>  $\langle$PAGES$\rangle$ 239-290 $\langle$/PAGES$\rangle$  \\
         6. \> \> \> \>  $\langle$TITLE$\rangle$ Computer Driven Displays and Their Use in Man/Machine Interaction. $\langle$/TITLE$\rangle$  \\
         7. \> \> \> \>  $\langle$TYPE$\rangle$ article $\langle$/TYPE$\rangle$  \\
         8. \> \> \> \>  $\langle$URL$\rangle$ http://dblp.uni-trier.de/db/journals/ac/ac7.html$\#$Dam66 $\langle$/URL$\rangle$  \\
         9. \> \> \> \>  $\langle$VOLUME$\rangle$ 7 $\langle$/VOLUME$\rangle$  \\
         10. \> \> \> \>  $\langle$YEAR$\rangle$ 1966 $\langle$/YEAR$\rangle$  \\
				 11. \> \> \> $\langle$row$\rangle$ \\
        12. \> \>  $\langle$/PUBLICATION$\rangle$  \\
      \end{tabbing}
    \end{rm}
  	\caption{\label{fig:dblp-xml}	Example XML entry extracted from the DBLP Schema.}
  \end{center}            
\end{figure*}

Attribute names are also indexed as individual keywords so that a query ``Anatomy
of a search engine author'' should return trails from the Anatomy paper to the
entries for Sergey Brin and Larry Page. XML documents discovered on web sites are
automatically recognized as such and can be indexed in the same way, as can
XML documents stored in the database, thus increasing coverage.

\section{Query Expressiveness}
\label{sec:query}

We have extended the search engine style query syntax to support an attribute
container operation using the ``='' sign. The construct $x=y$ means that an
attribute $y$ must be contained in an XML tag $x$. For example, the query
``Simon'' might return publications relating to Simon's probabilistic model
as well as articles by authors named Simon. The query \texttt{author=simon}
would restrict the returned entries to those contained in an XML attribute
$\langle$\texttt{author}$\rangle$, which translates to those in the author
table. i.e. publications written by authors named Simon. This is achieved by
indexing attribute, value pairs in the inverted file. The approach is expensive
in its use of disk space but retains fast access. The search engine
query operations such as \texttt{+}, \texttt{-} and \texttt{link:}
still remain supported with this extension. Thus a query
``Computers -type=phdthesis -type=mastersthesis'' would return books, journals
and articles on Computers, but no theses. This syntax does require some
knowledge of either table or attribute names, but exists as an option to allow
those with such knowledge to gain greater control.

This means we can provide trails which answer disjunctive queries (the default),
with preference for results containing as many keywords as possible (conjunctive).
We can also force the return of trails containing only specific keywords or which
exclude certain keywords. We can also use the attribute syntax to provide more
complex selection. For example, the query ``Computers -type=phdthesis -type=mastersthesis''
would be equivalent (using the DBLP webcase) to the query

\begin{tabbing}
t1\=t2\=t3\=t4\=t5\=t6\=t7\=t8\=t9\=t10\=t11\=t12\= \kill \\
  \> \texttt{select} * \texttt{from} publication \\
	\> \> \texttt{where} title \texttt{like} '$\%$Computers$\%$' \\
  \> \> \texttt{where} type $<>$ 'phdthesis' \\
  \> \> \texttt{and} type $<>$ 'mastersthesis' \\
\end{tabbing}

This is not a major saving. However, a researcher who is trying to find 
the year of publication of Brin and Page's search engine paper \cite{BRIN98}
could find the answer with  a query such as ``sergey anatomy'', whereas the
full SQL required would be:

\begin{tabbing}
t1\=t2\=t3\=t4\=t5\=t6\=t7\=t8\=t9\=t10\=t11\=t12\= \kill \\
  \> \texttt{select} year \texttt{from} publication, writes, author \\
	\> \> \texttt{where} lower(author.name) \texttt{like} 'sergey$\%$' \\
	\> \> \texttt{and} lower(publication.title) \texttt{like} 'anatomy$\%$' \\
  \> \> \texttt{and} writes.publication = publication.key \\
  \> \> \texttt{and} writes.author = author.id; \\
\end{tabbing}

We believe the DbSurfer expression represents a significant saving in
time and complexity for the user, whilst still returning the desired result.
Using Oracle's \texttt{explain plan} function \cite{GURR96} to examine the
actions of the Oracle database when performing this query reveals that 8
operations are required to complete this query including a full table scan.
Other relational databases are likely to offer similar performance. In
comparison, the DbSurfer results require no database accesses to compute
the trails, and require only 3 index-only accesses to examine the relevant
entries, showing that DbSurfer can provide results which provide savings
in database activity as well as user input.

\section{Examples}
\label{sec:examples}

\begin{figure*}[htbp]
  \begin{center}
		\psfig{figure=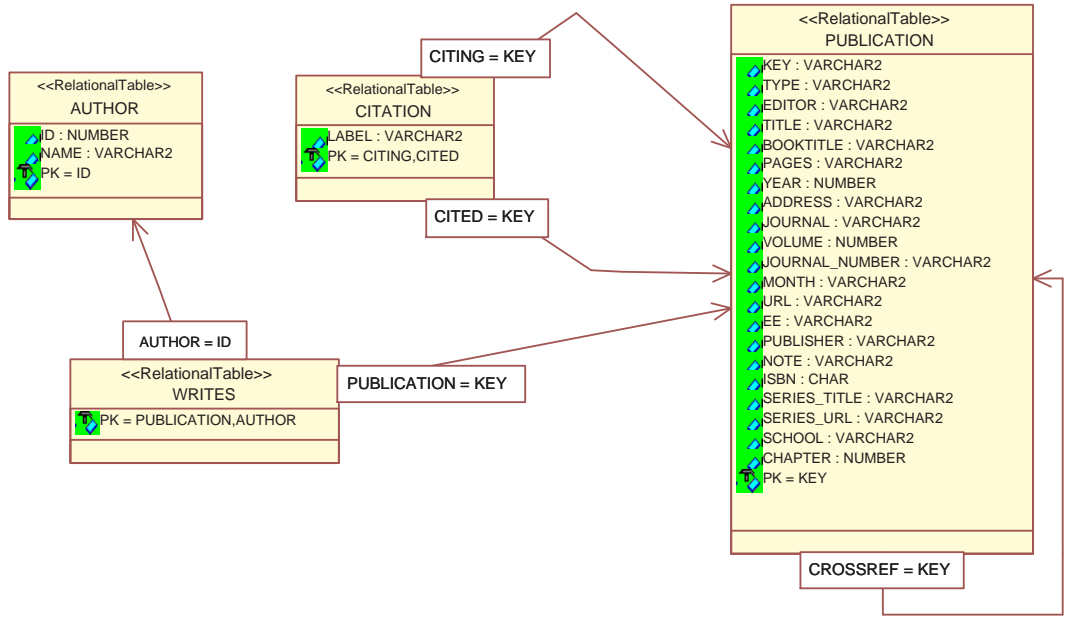, width=450pt}
  	\caption{
			\label{fig:dblp-schema}UML Diagram showing the DBLP Schema.
			The publication table stores details of all the journal, article
			and book entries, indexed by the attribute \texttt{key}.
			The \texttt{citation} table refers to two publication entries
			using the foreign keys \texttt{cited} and \texttt{citing}. Finally,
			The \texttt{Author} table is indexed on the primary key \texttt{id},
			and is linked to the \texttt{publication} table by the writes table,
			whose foreign keys are \texttt{publication}, which refers to the \texttt{key}
			attribute in the \texttt{publication} table and \texttt{author} which
			refers to the \texttt{id} field in the \texttt{author} table.
		}
  \end{center}            
\end{figure*}

In order to highlight the differences between the varying keyword-based systems 
for indexing relational database content, we have followed Hulgeri's lead in
indexing the content of a relational database containing DBLP data
\cite{HULG01} \cite{LEY02}. The DBLP
data is downloaded as an XML file which we then parsed to create the schema
shown in figure~\ref{fig:dblp-schema}. There are four tables in the schema.
The \texttt{publication} table (230000 rows, 300Mb) holds details of all the
journal, article and book entries. The \texttt{author} table
(150000 rows, 20Mb) contains details of each individual author, and the
\texttt{writes} table (480000 rows, 20Mb) links these together. The
\texttt{citation} table (100000 rows, 13Mb) links publications with those
which reference them.

\begin{figure*}[htbp]
  \begin{center}
		\psfig{figure=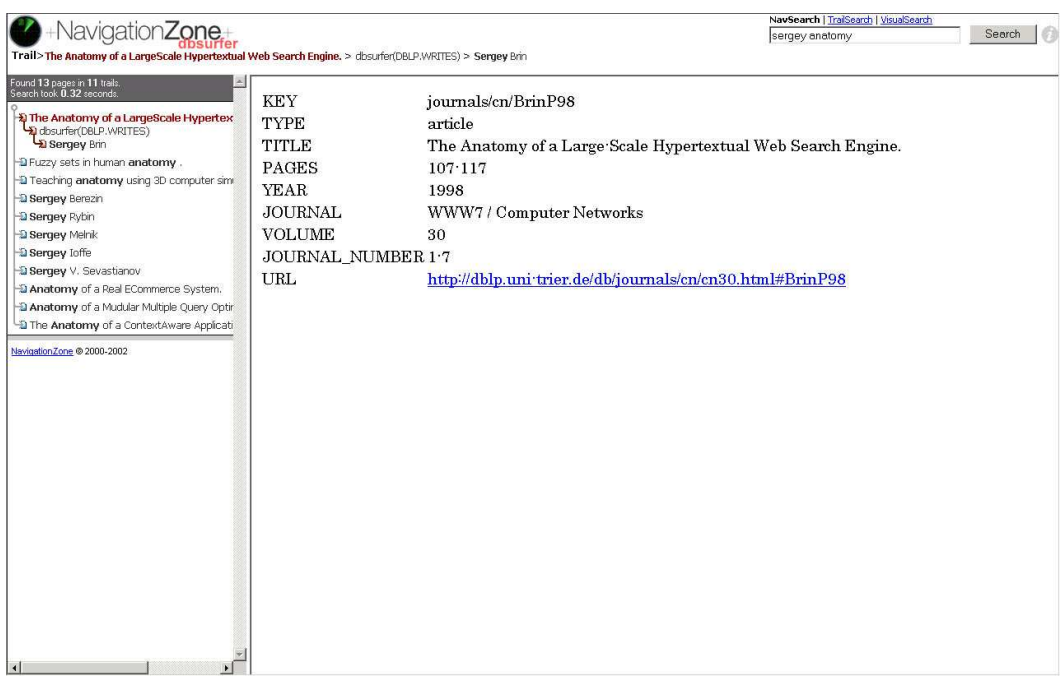, width=140mm}
  	\caption{ \label{fig:SergeyAnatomy}
			Example results using DbSurfer for the query ``sergey anatomy''.
		}
  \end{center}            
\end{figure*}

\begin{figure*}[htbp]
  \begin{center}
		\psfig{figure=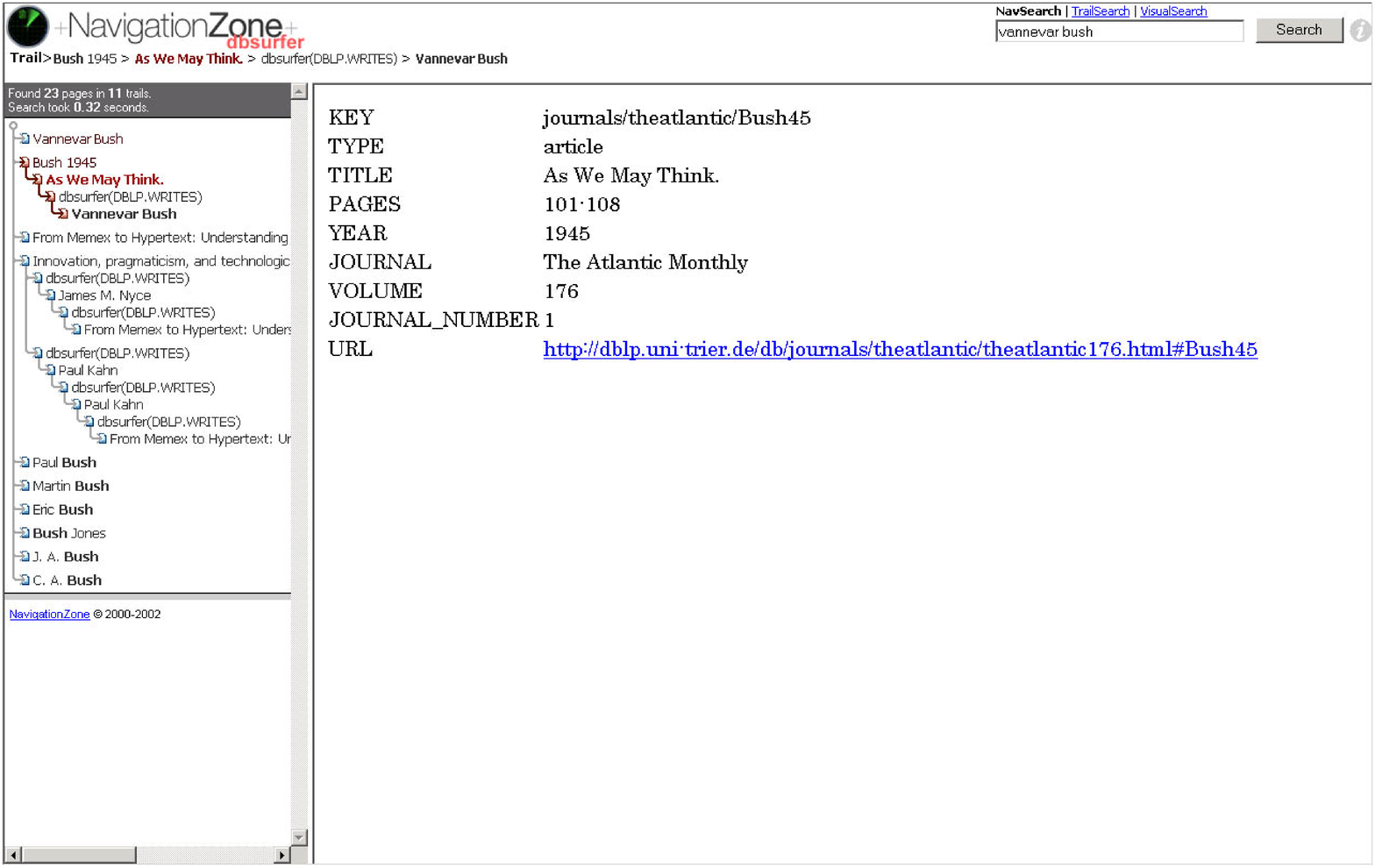, width=140mm}
  	\caption{ \label{fig:VannevarBush}
			Example results using DbSurfer for the query ``vannevar bush''.
		}
  \end{center}            
\end{figure*}

We have made the DBLP interface available to the public as a demonstration of
DbSurfer's potential. This demo can be reached from the homepage for Birkbeck
College School of Computer Science's Web Navigation Group at
\href{http://nzone.dcs.bbk.ac.uk/}{http://nzone.dcs.bbk.ac.uk/}.
Figure~\ref{fig:SergeyAnatomy} and figure~\ref{fig:VannevarBush} show two
examples of the NavSearch interface used for both database and web search.

Figure~\ref{fig:SergeyAnatomy} shows results for the query ``sergey anatomy''.  The first trail
shows the entries for Sergey Brin and Brin and Page's much-cited paper ``Anatomy of
a Large Scale Hypertextual Web Search Engine'' \cite{BRIN98}. In this example, the
remaining trails are single-node trails describing other authors called Sergey and
other papers with anatomy in the title.

Figure~\ref{fig:VannevarBush} shows results for the query ``vannevar bush''. The first
trail is a singleton node showing the author entry for Vannevar Bush. The second shows
Bush's paper ``As we may think'' \cite{BUSH45} in the context of a citation by a later
work. The third trail shows two papers describing work related to Vannevar Bush and
Memex, both by James M. Nyce.

It should be noted that the DBLP already has a search system designed
specifically for researcher's needs. The DbSurfer system cannot hope to replace
all the functionality of a custom system or of a relational database. The reason
for choosing the DBLP as a demonstration is to allow better testing and comparison
with similar databases-indexing systems. However, DbSurfer would allow the rapid
deployment of a search and navigation interface in situation where no such
interface exists. Secondly, DbSurfer can allow the development of a custom system
by using XSLT stylesheets to format results. In many cases, missing features and
aggregation of results can be added by constructing views at the database level.

\section{Evaluation}
\label{sec:evaluation}

As a preliminary evaluation into the relative performance of DbSurfer, we ran
two experiments. These were performed on a server with 1GHz dual Pentium III
processors.

In the first experiment, we selected the 20 papers found in the DBLP corpus,
with the highest ranks in the ResearchIndex (CiteSeer) ``most accessed documents''
list. From this we constructed 20 queries by taking the surname of the first
author and 1, 2 or 3 significant keywords with which a user might expect to identify
that paper. We submitted these queries to DbSurfer for evaluation. We also submitted
them to compared BANKS (Browsing ANd Keyword Search in relational databases)
\cite{HULG01} and CiteSeer \cite{LAWR99} for comparison. The results are shown in
figure~\ref{fig:eval}. The key result is that
DbSurfer performs well (and outperforms BANKS and Citeseer) in finding requested
references. The table shows reciprocal ranks for the desired paper,
in terms of the trail, page or cluster containing the relevant citation. Only
the first page of results was considered in each case, but this should have minimal
impact on the results. 
Times are shown as reported by each of the systems concerned and are not strictly
comparable, but are intended to be indicitive of the general level of performance.
Times are also missing for those queries for which the BANKS system failed to return
any results.

This result is encouraging, but may be misleading in places.
The poor retrieval performance of BANKS is largely due to its poor coverage
as it indexes only a subset of the DBLP data set. The 21.38 second response time
for the query ``nilsson routers'' is due to bad configuration and behaviour of
the garbage collector. However, a top-and-tailed average time of 1.2 seconds is
still disappointingly short of the sub-second response time expected. More worrying
is that a third of queries failed to return the desired document in any of the
returned trails. However, over half the desired documents where identified in
the best trail for each query, suggesting that the trail-finding scheme can be
highly effective.

\begin{figure*}[ht]
 \begin{center}
\begin{tabular}{|l|rr|rr|r|}
\hline
	& \multicolumn{2}{c}{DbSurfer} & \multicolumn{2}{c}{Banks} & Citeseer \\
Query & 1/Rank & Time & 1/Rank & Time & 1/Rank \\ \hline
crescenzi ip lookup                 & 0.00 &  0.40 & 0.00 & 11.77 & 0.13 \\
web database florescu               & 0.33 &  1.33 & 0.00 &       & 0.00 \\
brin anatomy                        & 1.00 &  0.35 & 0.00 &       & 0.00 \\
digital libraries lawrence          & 0.00 &  1.74 & 0.00 &       & 0.00 \\
waldvogel ip routing                & 1.00 &  0.77 & 0.00 &       & 0.33 \\
rivest cryptosystems                & 1.00 &  1.22 & 1.00 &  2.93 & 0.25 \\
web mining cooley                   & 0.00 &  1.59 & 0.00 &       & 1.00 \\
broch routing                       & 1.00 &  1.43 & 0.00 &  0.93 & 0.06 \\
deerwester latent semantic analysis & 1.00 &  1.13 & 0.00 & 12.27 & 0.20 \\
agrawal mining                      & 0.33 &  2.20 & 0.00 &       & 0.00 \\
bryant boolean function             & 1.00 &  0.70 & 0.00 &       & 0.00 \\
nilsson routers                     & 0.00 & 21.38 & 0.00 &  0.93 & 1.00 \\
rcs tichy                           & 1.00 &  0.93 & 0.00 &  1.32 & 1.00 \\
traffic leland                      & 1.00 &  0.99 & 0.00 &       & 0.00 \\
joachims support vector             & 0.00 &  1.17 & 0.00 & 10.27 & 0.06 \\
traffic paxson                      & 1.00 &  0.69 & 0.00 &       & 0.00 \\
time elman                          & 0.00 &  1.78 & 0.00 &  1.84 & 0.00 \\
workflow georgakopoulos             & 1.00 &  1.60 & 0.00 &       & 1.00 \\
ferragina b-tree                    & 1.00 &  1.31 & 0.00 & 13.18 & 0.20 \\
fraley clusters                     & 0.00 &  0.72 & 0.00 &       & 0.00 \\ \hline
Average	                            & 0.58 &  2.17 & 0.05 &  6.16 & 0.26 \\ \hline
\end{tabular}
    \caption{
			\label{fig:eval}
			Comparison of reciprocal rank and total time taken for 20 citation-seeking
			queries on DbSurfer, BANKS and CiteSeer.
		}
\end{center}
\end{figure*}

The second experiment provided a closer analysis of the times taken is computing
the results. By isolating two papers and requesting them with a decreasing number
of keywords, we could analyse the times taken to perform each component operation.
Computing scores for nodes takes around 50\% of the total processing time, with
the trail finding taking around 30\%, computing the text summaries around 15\%,
filtering redundant information around 2\%, with the remainder being taken up by
system overhead, XML transformation and presentation. Increasing the number of
keywords causes a limited increase in the time to compute page scores, but this
impact is dwarfed by other factors. 
One other interesting result is that as the number of keywords increases so does
the fraction of nodes in the returned trails which are distinct for the entire
trailset. Only extensive user testing will confirm whether this is a positive
feature.



\section{Related Work}
\label{sec:related}

Recent work at Microsoft Research, at the Indian Institute of Technology, Bombay
and at the University of California has resulted in several systems similar in
many ways to our own. However, the system we describe differs greatly in the
design of the algorithms and in the style of the
returned results. Our system also offers the opportunity for
integrating both web site and database content with a common interface and for
searching both transparently.

BANKS was developed by the Indian Institute of Technology \cite{HULG01}. Each result
in the BANKS systems is
a tree from a selected node, ordered by a relevance function which factors in node
and link weights. Mragyati, also developed at the Indian Institute of Technology,
uses a similar approach in which keyword queries are converted to SQL at query time
\cite{SARD01}. This approach has some notable advantages. It guarantees that all
data being searched on is fresh, whereas DbSurfer only ensures that the displayed
data is fresh - the data in the inverted file will need to be periodically updated
to ensure that it is not ``stale''. The authors claim that
the approach ``is scalable, as it does not build an in-memory graph''. This is
a legitimate criticism of DbSurfer's approach. However, allowing almost arbitary
selection of attributes for querying and relying on the databases own indexes
restricts the indexing of binary fields to those supported by the database (usually
in non-standard components) and makes full-table scans probable, introducing a new
problem in scalability and response time. In addition, research has shown that
large graphs (e.g. a few billion nodes) can be stored and manipulated in main memory of
mid-range servers when appropriate compression techniques are used\cite{BOLD03,RAND01}.

DBXplorer \cite{AGRA02} was developed by
Microsoft Research, and like BANKS and Mragyati, it uses join trees to compute an SQL
statement to access the
data. The algorithm to compute these differs, as does the implementation, which was
developed for Microsoft's IIS and SQL Server, the others being implemented in Java.
DbSurfer does not require access to the database to discover the trails, only to display
the data when user clicks on a link in that trail.

DISCOVER is the latest offering and
shares many similarities to Mragyati, BANKS and DbXplorer, but uses a greedy algorithm
to discover the \emph{minimal joining network} \cite{VAGE02}. It also takes greater
advantage of the database's internal keyword search facilities by using Oracle's Context
cartridge for the text indexing.

Goldman et al. have also introduced a system for keyword search \cite{GOLD98}. Their
system works by finding results for queries of the form $x$ \texttt{near} $y$ (e.g.
find movie near travolta cage). Two sets of entries are found - and the contents of
the first set are returned based upon their proximity to members of the second set.
In comparison to DbSurfer, there is no support for navigation of the database (manual
or assisted) nor any display of the context of the results.

The join discovery problem is related to the problem tackled by the
\emph{universal relation model} \cite{ULLM89} \cite{LEVE92}. The idea underlying
the universal relation model is to allow querying the database soley through its
attributes without explicitly specifying the join paths. The expressive querying
power of such a system is essentially that of a union of conjunctive queries
(see \cite{SAGI83}). DbSurfer takes this approach further by allowing the user to
specify values (keywords) without stating their related attributes and providing
relevance based filtering.

Goldman and Widom outline an approach for the related problem of
allowing structured database queries on the web \cite{GOLD00}.
WSQ/DSQ (pronounced ``wisk-disk'')
is a combination of two systems for Web-Supported and Database-Supported
Queries. WSQ allows structued queries on web data, by allowing two virtual
tables, $WebPages(SearchExp$, $T_1$, $T_2 \ldots T_n$, $URL Rank$, $Date)$ and
$WebCount(SearchExp$, $T_1$, $T_2 \ldots T_n$, $Count)$, both of which can be
queried alongside normal RDBMS tables. A similar approach to \cite{GOLD00} is
adopted by Squeal, which provides \texttt{page}, \texttt{tag}, \texttt{att},
\texttt{link} and \texttt{parse} tables which can be queried using
SQL \cite{SPER00}. It would be possible to extend the DbSurfer engine to
provide such functionality by adding appropriate stored procedures to the database.
These could request data from DbSurfer (using SOAP or a similar RPC protocol) and
map the returned trail information to an appropriate schema, such as that described
by Heather and Rossiter \cite{HEAT90}.

\section{Future Work}
\label{sec:future}

In addition to improving the quality of the overall results and speed of delivery,
the issue of incremental updates needs to be addressed. Theoretically, this could
be achieved by storing a simple checksum of the database field values alongside
the main index. The database could then be queried for those rows where the
checksum is different. This restriction could be added at the database level, or
in DbSurfer prior to construction of the virtual document. When this work is
finished, several other key problems will still remain.

\subsection{Queries}

The current system does not handle range queries. In fact DbSurfer does not
handle numbers very well - it works only using the text representation. We can improve this
situation by following ideas presented in \cite{AGRA02b}. The system described
recognizes numbers in both documents and queries and looks for close matches.
The same strategy can be extended to dates, by converting all date representations
to numeric values. 
Given attribute-value pairs in the inverted file, we can implement some aggregate
functions by combining values at query time. An alternative strategy is to index
views created at the database level, but this requires a good understanding of the
values which are likely to be aggregated.

The evaluation of the system is encouraging, but limited. In order to achieve a
more comprehensive comparison, we propose the creation of an independant test
suite for database keyword-search, with a competition run on similar lines to the
TREC conference, perhaps as a workshop associated with a major conference.

\subsection{Presentation}

Some presentation issues exist for the row display servlet.
Backlink handling, for example, is an issue. When navigating the database structure
it should be possible to examine those rows which reference a given attribute. This
can be achieved by using a separate servlet to generate the list of rows, which might
operate by submitting another query with the command \texttt{link:currenturl}. This
would return a list of rows which reference the current page's underlying row. With
the appropriate query modification, this could be extended to restrict entries to
the user's requirements.

Another issue is the handling of multipart keys. Each foreign key field is currently
displayed as an outlink. However, this method of display does not extend to multipart
or composite keys. In particular, it will not work for composite keys where one of
the component attributes is a foreign key for some other table. In such a situation
it is unclear were the destination of such a link should be.

\subsection{Security}

Security is a major issue. By constructing a single index we remove the
fine-grained access controls employed by the database. Since all indexing is done
through a single user account, the access rights for all DbSurfer users are
equivalent to the access rights of that user. One possible way to restore some of
the fine-grained security may be to allow each user to view the data only
under the database username and password which they supply.	Such a system might be
implemented using container managed security which is part of the J2EE standard.
This would require some very simple server configuration and a view on the data
dictionary tables of the underlying RDBMS. However, this is not a complete solution
as it would only affect the display servlet. We would need to expand this so that
rows which could not be displayed were never presented to the user. This would have
a noticable impact on performance. However, failure to do this would have two
negative implications. Firstly, the system would present users with data which they could
not access (this being analogous to returning
404s in a web search engine). Secondly, it might be possible to infer information
without the rows being displayed. For example, if a company had an invoices table
indexed, simply the presence of an entry (for example \texttt{payee=enron} or
\texttt{reason=takeover}) might be considered damaging. Until these issues are
resolved, the efficient indexing of secure data for unstructured search will be
highly problematic.

\section{Concluding Remarks}

We have presented DbSurfer - a system for keyword search and navigation through
relational databases. DbSurfer's unique feature is a novel join discovery
algorithm which discovers Memex-like trails though the graph of foreign-to-primary
key dependencies. DbSurfer allows queries to be answered efficiently, providing
relevant results without relying on a translation to SQL.

\bibliographystyle{plain}
\bibliography{../bib/papers}

\end{document}